\documentclass[12pt,aps,prb,preprint]{revtex4}
\begin{document}
\newcommand{\br}{{\bf r}}
\title{Green's functions for Neumann boundary conditions}
\author{Jerrold Franklin}
\email{Jerry.F@TEMPLE.EDU}
\affiliation{Department of Physics,
Temple University, Philadelphia, PA 19122-6082}
\date{\today}
\begin{abstract}
Green's functions for Neumann boundary conditions have been considered in Math Physics and Electromagnetism textbooks, but special constraints and other properties required for Neumann boundary conditions have generally not been noticed or treated correctly.   In this paper, we derive an appropriate Neumann Green's function with these constraints and properties incorporated.
\end{abstract}
\maketitle

\section{Introduction}
The Green's function method for solving Sturm-Liouville problems of the form
\begin{equation}
{\cal L}_{\br}\phi(\br)=\nabla\cdot[p(\br)\nabla\phi(\br)]+q(\br)\phi(\br)=\rho(\br)
\label{eq:sl}
\end{equation}
 is described in many textbooks. 
We list some of the more recent texts that treat Green's functions in references 1-4,
and review the usual textbook treatment below. 
For simplicity, we will take $q(\br)=0$.
%Our general conclusions also apply to the case with $q(\br)$, but some of the equations would be more complicated.

A motivation for using a Green's function, $G(\br,\br')$, is that it satisfies homogeneous boundary conditions that makes it easier to solve for than the original problem with inhomogeneous boundary conditions.  Application of Green's theorem
\begin{equation}
\int[\phi(\br'){\cal L}_{\br' }G(\br,\br')-G(\br,\br'){\cal L}_{\br'}\phi(\br')]d\tau' =
\int{\bf dS'}\cdot[\phi(\br')p(\br')\nabla' G(\br,\br')-G(\br,\br')p(\br')\nabla'\phi(\br')]
\label{eq:gt}
\end{equation}
 provides the solution to the original problem if $G({\bf r,r'})$ satisfies
\begin{equation}
{\cal L}_{\br'}G({\bf r,r'})=\delta({\bf r-r'}).
\label{eq:delta}
\end{equation}
In that case, Eq (\ref{eq:gt}) leads to 
\begin{equation}
\phi(\br)=\int G(\br,\br')\rho(\br')d\tau'+\int{\bf dS'}\cdot[\phi(\br')p(\br')\nabla' G(\br,\br')-G(\br,\br')p(\br')\nabla'\phi(\br')].
\label{eq:gts}
\end{equation} 

The two simplest boundary conditions for which the Green's function method is applicable are the Dirichlet boundary condition for which the solution $\phi(\br)$ is given
 on all bounding surfaces, and the Neumann boundary condition for which its normal derivative ${\bf\hat n}\cdot\nabla\phi(\br)$ is given. 
The Dirichlet Green's function is generally used for electrostatic problems where the potential is specified on bounding surfaces, while
the Neumann Green's function is useful for finding temperature distributions where the bounding surfaces are heat insulated or have specified heat currents.

For the Dirichlet boundary condition, the Green's function satisfies the homogeneous boundary condition
\begin{equation}
G_D(\br,\br')=0,\quad{\rm for}\;\br'\;{\rm on\;all\;bounding\;surfaces.}
\label{eq:dbc}
\end{equation}
This reduces Eq. (\ref{eq:gts}) to
\begin{equation}
\phi(\br)=\int G_D(\br,\br')\rho(\br')]d\tau'+\int{\bf dS'}\cdot[\phi(\br')p(\br')\nabla' G_D(\br,\br')],
\label{eq:dsol}
\end{equation}
which is the solution to the Dirichlet problem once $G_D(\br,\br')$ is known.
The Dirichlet Green's function is the solution to Eq.\ (\ref{eq:delta}) satisfying the homogeneous boundary condition in Eq.\ (\ref{eq:dbc}).

Textbooks generally treat the Dirichlet case as above, but do much less with the Green's function for the Neumann boundary condition, and what is said about the Neumann case often has mistakes of omission and commission.  First of all, the Neumann boundary condition for the solution $\phi(\br)$ must satisfy  the constraint
\begin{equation}
\int{\bf dS}\cdot p(\br)\nabla\phi(\br)=\int\rho d\tau,
\label{eq:con}
\end{equation}
which follows from applying the divergence theorem to Eq.\ (\ref{eq:sl}).  Most texts do not mention this important constraint on the Neumann boundary 
condition.\cite{con}

There are cases where the boundary condition is Neumann on some surfaces and Dirichlet on others.  In those cases, the normal derivative of $\phi$ on  the Dirichlet surfaces automatically adjusts to satisfy the constraint surface integral.  But for pure Neumann boundary conditions, the normal derivative must satisfy the constraint or no solution exists.  In the following sections, we will assume pure Neumann boundary conditions for which constraint equation (\ref{eq:con}) holds.  We treat the one dimensional Neumann Green's function in Section 2, and then the three dimensional case in Section 3.

\section{1D Neumann Green's function}

We first review how the Dirac delta function arises when a function $f(x)$, defined in the finite range  $0\le x\le L$,
is expanded in orthonormal  Dirichlet eigenfunctions $u_n(x)$ of a Sturm-Liouville operator ${\cal L}_x$.
The eigenfunctions satisfy the differential equation
\begin{equation}
 {\cal L}_x u_n(x)=\frac{d}{dx}\left[p(x)\frac{du_n}{dx}\right]=\lambda_ n u_n(x).
\label{eq:ldf}
\end{equation}
with homogeneous Dirichlet boundary conditions  
\begin{equation}
u_n(0)=0,\quad u_n(L)=0.
\label{eq:df}
\end{equation}

The function $f(x)$ can be expanded in the Dirichlet eigenfunctions as
\begin{equation}
f(x)=\sum_{n=1}^\infty b_n u_n(x),
\label{eq:dex}
\end{equation}
with the expansion coefficients $b_n$ given by
\begin{equation}
b_n=\int_0^L u_n^*(x) f(x) dx.
\label{eq:bn}
\end{equation}
If Eq.\ (\ref{eq:bn}) for the expansion coefficients is substituted into Eq.\ (\ref{eq:dex}) for $f(x)$, and the sum executed before the integral, we get  
\begin{equation}
f(x)=\int_0^L dx'
\sum_{n=1}^\infty  u_n^*(x')  u_n(x)f(x').
\label{eq:ff'}
\end{equation}

From the definition of the Dirac delta function by its sifting property,
\begin{equation}
f(x)=\int_0^L dx'
\delta(x-x')f(x')
\label{eq:def}
\end{equation}
for $x$ and $x'$ in the range $[0,L]$,
we see that the delta function can be represented by a sum over Dirichlet eigenfunctions as
\begin{equation}
\delta(x-x')=
\sum_{n=1}^\infty  u_n^*(x')  u_n(x).
\label{eq:dsum}
\end{equation}
For example, for the simple case
\begin{equation}
 {\cal L}_x u_n=u_n^{''}=\lambda_n u_n,\quad u_n=\sqrt{\\2/L}\sin(n\pi x/L),\quad \lambda_n=-\left(\frac{n\pi}{L}\right)^2,
\label{eq:sin}
\end{equation}
the delta function is represented by
\begin{equation}
\delta(x-x')=(2/L)
\sum_{n=1}^\infty \sin(n\pi x'/L)\sin(n\pi x/L)
\label{eq:dsin}
\end{equation}

A Dirichlet Green's function that satisfies the differential equation
\begin{equation}
{\cal L}_{x'}G_D(x,x')=\delta(x-x'),
\label{eq:lgd}
\end{equation}
and satisfies the homogeneous Dirichlet boundary conditions in the variable $x'$ can be formed from the Dirichlet eigenfunctions as
\begin{equation}
G_D(x,x')=
\sum_{n=1}^\infty\frac{u^*_n(x') u_n(x)}{\lambda_n}.
\label{eq:dgs}
\end{equation}
Acting on this Green's function with the Sturm Liouville operator ${\cal L}_{x'}$ removes the denominator in Eq.\ (\ref{eq:dgs}) leaving the delta function of 
Eq.\ (\ref{eq:lgd}), showing that  this is the appropriate Dirichlet Green's function.

The above straightforward derivation for Dirichlet boundary conditions is given in most texts, but a corresponding derivation for Neumann boundary  conditions is generally absent.   Neumann eigenfunctions, $v_n(x)$, satisfy the same differential equation  (\ref{eq:ldf}) as the Dirichlet eigenfunctions, but have the boundary conditions
\begin{equation}
v_n'(0)=0,\quad v'_n(L)=0.
\label{eq:vbc}
\end{equation}

The expansion in Neumann eigenfunctions has a constant eigenfunction corresponding to a zero eigenvalue, 
so the expansion is given by 
\begin{equation}
f(x)=a_0+\sum_{n=1}^\infty a_n v_n(x),
\label{eq:nex}
\end{equation}
The Neumann expansion coefficients $a_n$ are given by the integrals
\begin{equation}
a_n=\int_0^L v_n^*(x) f(x) dx,\quad n\ge 1,
\label{eq:an}
\end{equation}
and
\begin{equation}
a_0=\frac{1}{L}\int_0^L f(x) dx=<f>,
\label{eq:az}
\end{equation}
where $<f>$ represents the average value of the function $f(x)$ over the interval $[0,L]$.
Now putting Eqs.\ (\ref{eq:an}) and (\ref{eq:az}) into the expansion Eq.\ (\ref{eq:nex}) results in
\begin{equation}
f(x)=\int_0^L dx'\left[\frac{1}{L}+
\sum_{n=1}^\infty  v_n^*(x')  v_n(x)\right]f(x'),
\label{eq:nff'}
\end{equation}
so the representation of the delta function in terms of Neumann eigenfunctions is
\begin{equation}
\delta(x-x')=\frac{1}{L}+
\sum_{n=1}^\infty  v_n^*(x')  v_n(x).
\label{eq:nsum}
\end{equation}
The additional constant term $1/L$ is not generally recognized in textbooks.
For the simple case of ${\cal L}_x v_n=v''_n=\lambda_n v$ with the eigenfunctions satisfying the homogeneous Neumann boundary conditions 
of Eq.\ (\ref{eq:vbc})
the delta function is represented by
\begin{equation}
\delta(x-x')=\frac{1}{L}+
\frac{2}{L}\sum_{n=1}^\infty  \cos(n\pi x'/L)\cos(n\pi x/L).
\label{eq:nc}
\end{equation}

A Neumann's Green function can be formed using Neumann eigenfunctions of the operator ${\cal L}_{x}$ as the sum
\begin{equation}
G_N(x,x')=
\sum_{n=1}^\infty\frac{v_n^*(x')  v_n(x)}{\lambda_n}.
\label{eq:ngs}
\end{equation}
This Green's function satisfies the homogeneous Neumann boundary conditions
\begin{equation}
\partial_{x'} G_N(x,x')|_{ (x'=0)}=0,\quad \partial_{x'}G_N(x,x')|_{(x'=L)}=0.
\label{eq:ngbc}
\end{equation}
However, because of the constant term $1/L$ in Eq.\ (\ref{eq:nsum}), 
the operation on $G_N(x,x')$ by ${\cal L}_{x'}$ is
\begin{equation}
{\cal L}_{x'}G_N(x,x')=\delta(x-x')-1/L.
\label{eq:lgn}
\end{equation}
Thus, the Neumann Green's function satisfies a different differential equation than the Dirichlet Green's function.

We now use the Green's function $G_N(x,x')$ to find the solution of the differential equation
\begin{equation}
{\cal L}_x f(x)=
\frac{d}{dx}\left[p(x)\frac{df}{dx}\right]=\rho(x),
\label{eq:ndf}
\end{equation}
with the inhomogeneous Neumann boundary conditions
\begin{equation}
f'(0)=f'_0,\quad f'(L)=f'_L.
\label{eq:nfbc}
\end{equation}
The boundary values must satisfy the  constraint 
\begin{equation}
p(L)f'_L-p(0)f'_0=\int_0^L\rho(x)dx,
\label{eq:const}
\end{equation}
which follows from a first integral of Eq.\ (\ref{eq:ndf}).
The Neumann Green's function must also satisfy this constraint, which it does because the right hand side of Eq.~({\ref{eq:lgn}) integrates to zero.

Green's theorem in one dimension (or integration by parts) for this differential equation leads  to
\begin{equation}
\int_0^L [f(x'){\cal L}_{x'}G_N(x,x')-G_N(x,x'){\cal L}_{x'}f(x')]dx'=
-G_N(x,L)p(L)f'_L+G_N(x,0)p(0)f'_0.
\label{eq:ngth}
\end{equation}
We have used the boundary conditions (\ref{eq:ngbc}) to eliminate terms containing $\partial_{x'}G(x,x')$ at the endpoints.
Then, using Eq.\  (\ref{eq:lgn}), we get
\begin{equation}
f(x)=<f>+\int_0^L G_N(x,x')\rho(x')dx'-G_N(x,L)p(L)f'_L+G_N(x,0)p(0)f'_0,
\label{eq:sltn}
\end{equation}
which constitutes the solution to the Sturm-Liouville problem for Neumann boundary conditions.
The constant $<f>$, the average value of $f(x)$, arises when the term $1/L$ in Eq.\ (\ref{eq:lgn}) is substituted into Eq.\ (\ref{eq:ngth}). 

Actually, any constant can be added to the solution $f(x)$ since the solution of the Neumann problem is only unique up to an additive constant.
The constant term in Eq.\ (\ref{eq:sltn}) will always be the average value of the solution because the variable terms have zero average value.

Although we have used an expansion in eigenfunctions to give an heuristic derivation of the Green's function for Neumann boundary conditions, any function satisfying the defining equation (\ref{eq:lgn}) with the boundary conditions of Eq.\ (\ref{eq:ngbc}) will be a suitable Green's function $G_N (x.x')$.
For example, a Green's function for the problem
\begin{equation}
f''(x)=x,\quad f'(0)=f'_0,\quad f'(L)=f'_L=f'_0+L^2/2,
\label{eq:fpx}
\end{equation}
is given by
\begin{eqnarray}
0\le x'\le x:&G_{N1}(x,x')&=-\frac{x'^2}{2L}+x\nonumber\\
x\le x'\le L:&G_{N2}(x,x')&=-\frac{x'^2}{2L}+x'.
\label{eq:gfn}
\end{eqnarray}

The term $-x'^2/2L$ in $G_{N1}$ and $G_{N2}$ provides the $-1/L$ term in $\partial^2_{x'}G_N(x,x')$, and also  satisfies the homogeneous Neumann boundary condition at $x'=0$.  The $x'$ term in $G_{N2}$ satisfies the homogeneous Neumann boundary condition at $x'=L$, and also provides a unit step in $G_N$ at $x'=x$ so that the next derivative will give the delta function in $\partial^2_{x'}G_N(x,x')$.  Finally, the term $x$ in $G_{N1}$ makes $G_N(x,x')$ continuous at $x'=x$.  Putting this Green's function into Eq.\ (\ref{eq:sltn}) gives the solution to Eq.\ (\ref{eq:fpx}).

We note that the Green's function in Eq.\ (\ref{eq:gfn}) is not symmetric with respect to $x$ and $x'$.  To show that a Neumann Green's function need not be symmetric, we repeat the usual proof of symmetry here.  Applying Green's theorem to two Green's functions, $G(x_1,x')$ and $G(x_2,x')$, of a Sturm-Liouville operator $\cal L$ gives
 \begin{eqnarray}
&&\int_0^L [G(x_1,x'){\cal L}_{x'}G(x_2,x')-G(x_2,x'){\cal L}_{x'}G(x_1,x')]\nonumber\\
&&=\left[G(x_1,x')p(x')\partial_{x'}G(x_2,x')-G(x_2,x')p(x')\partial_{x'}G(x_1,x')\right]^{x'=L}_{x'=0.}
\label{eq:sym}
\end{eqnarray}
The right hand side of Eq.\ (\ref{eq:sym}) vanishes for either Dirichlet or Neumann homogeneous boundary conditions.
For a Dirichlet Green's function, with ${\cal L}_{x'}G_D(x,x')=\delta(x-x')$,
Eq.\ (\ref{eq:sym}) reduces to
\begin{equation}
G_D(x_1,x_2) -G_D(x_2,x_1)=0,
\label{eq:dsym}
\end{equation}
so a Dirichlet Green's function must be symmetric.
However, for a Neumann Green's function, \mbox{${\cal L}_{x'}G_N(x,x')=-1/L+\delta(x-x')$,}
and Eq.\ (\ref{eq:sym}) reduces to
\begin{equation}
G_N(x_1,x_2) -G_N(x_2,x_1)=
\frac{1}{L}\int_0^L[G_N(x_1,x') -G_N(x_2,x')]dx',
\label{eq:nsym}
\end{equation}
so a Neumann Green's function is not required to be symmetric.

We do know however, from the eigenfunction expansion  in Eq.\ (\ref{eq:ngs}) that any real 
 Neumann Green's function can be made symmetric.  For instance, adding the term $-x^2/2L$ to the Green's function in Eq.\ (\ref{eq:gfn}) will make that Neumann Green's function symmetric without changing any of its actions.  There is no need to do this however, since the non-symmetric Green's function is simpler, and either form will solve the original differential equation.

\section{3D Neumann Green function}

In this section, we extend the one dimensional results of the previous section to three dimensions.
In three dimensions, we seek the solution of the differential equation
\begin{equation}
{\cal L}_{\br}\phi(\br)=\nabla\cdot[p(\br)\nabla\phi(\br)]=\rho(\br)
\label{eq:lp3d}
\end{equation}
with the inhomogeneous Neumann boundary conditions on any bounding surface
\begin{equation}
{\bf\hat n}\cdot\nabla\phi(\br)=f(\br_S),
\label{eq:bc3d}
\end{equation}
where ${\bf\hat n}$ is the outward normal vector to the surface and $f(\br_S)$ is an almost arbitrary function specified on all surfaces.
A solution to Eq.\ (\ref{eq:lp3d}) exists only if the boundary conditions satisfy the constraint
\begin{equation}
\oint{\bf dS}\cdot p(\br)\nabla\phi(\br)=\int\rho(\br)d\tau.
\label{eq:l3d}
\end{equation} 
This constraint follows by applying the divergence theorem to Eq.\ (\ref{eq:lp3d}). 

The Neumann Green's function for this problem
satisfies the differential equation
\begin{equation}
{\cal L}_{\br'}G_N(\br,\br')=\delta(\br-\br')-1/V,
\label{eq:g3d}
\end{equation}
with the homogeneous boundary condition
\begin{equation}
{\bf\hat n'}\cdot\nabla' G_N(\br,\br')=0
\label{eq:gbc3d}
\end{equation}
on all surfaces.  This Green's function automatically satisfies the constraint
\begin{equation}
\int{\bf dS'}\cdot p(\br')\nabla'G_N(\br,\br')=0,
\label{eq:c3d}
\end{equation} 
which follows by applying the divergence theorem to Eq.\ (\ref{eq:g3d}). 

The extra term $1/V$ in Eq.\ (\ref{eq:g3d}) is the 3D equivalent of the term $1/L$ in Eq.\ (\ref{eq:lgn}),
and arises due to the constant term in any expansion  using Neumann eigenfunctions.
The $1/V$ goes to zero for an infinite volume (the so called `exterior problem'), but is important if the volume considered is finite (the `interior problem').

The solution to Eq.\ (\ref{eq:lp3d}) is given by 
\begin{equation}
\phi(\br)=<\phi>+\int G(\br,\br')\rho(\br')d\tau'-\int{\bf dS'}\cdot G(\br,\br')p(\br')\nabla'\phi(\br'),
\label{eq:3dsltn}
\end{equation} 
which follows from Green's theorem, and Eqs.\ (\ref{eq:g3d}) and (\ref{eq:gbc3d}) for $G_N(\br,\br')$.
As in the 1D case, $<\phi>$ is an arbitrary constant that equals the average value of $\phi$ in the volume. 
It can be chosen to be zero to simplify the equation.

An alternate method to develop a Neumann Green's function has been proposed in Ref.~4.
That method keeps Eq.~(\ref{eq:delta}) for the action of $\cal L_{\br'}$ on the Greens function.
The normal derivative of the Green's function must then satisfy the constraint 
\begin{equation}
{p(\br')\bf\hat n}\cdot\nabla' G(\br,\br')=1/S
\label{eq:bcj}
\end{equation}
on all bounding surfaces, where the constant $S$ is the total area of the bounding surfaces. 
This constraint follows from applying the divergence theorem to Eq.~(\ref{eq:delta}).

This method does not always work.  Kim and Jackson\cite{kj} have applied it to the case of two concentric spheres, where spherical symmetry allows the constant 
normal derivative in Eq.~(\ref{eq:bcj}), while also satisfying Eq.~(\ref{eq:delta}).  However, for more general geometries, such as for the temperature distribution inside a rectangular parallelepiped with insulated walls, there is no function that satisfies Eq.~(\ref{eq:bcj}), while also satisfying Eq.~(\ref{eq:delta}).  To satisfy Eq.~(\ref{eq:bcj})
at both the left and right hand face of the parallelepiped would require either a periodic function or a function with a discontinuous derivative, neither of which would be compatible with Eq.~(\ref{eq:delta}).  On the other hand, Eq.~(\ref{eq:ngs}) provides a suitable Neumann Green's function for the parallelepiped, or for any geometry that allows a complete set of Neumann eigenfunctions.
This alternate method would also not work for any one dimensional problem, because it would require the first and second derivatives of $G(\br,\br')$ to each be of dimension 
1/length, which is not possible.

\section{Summary}

To summarize, we have found seven essential differences between the Neumann Green's function treated here and the Dirichlet Green's function which is generally treated in Math Physics  or Electromagnetism texts for the solution of a partial differential equation of the form 
\begin{equation}
{\cal L}_{\br}\phi(\br)=\nabla\cdot[p(\br)\nabla(\br)]\phi(\br)=\rho(\br).
\label{eq:lp3ds}
\end{equation}
We list again the different Neumann properties below:
\\ \\
\noindent
(1)  The Neumann boundary condition for the solution $\phi(\br)$ is
\begin{equation}
{\bf\hat n}\cdot\nabla\phi(\br)=f(\br_S),
\label{eq:bc3ds}
\end{equation}
(2)  with the constraint
\begin{equation}
\int{\bf dS}\cdot p(\br)\nabla\phi(\br)=\oint\rho(\br)d\tau
\label{eq:l3ds}
\end{equation} 
required for any solution to exist.
\\ \\
\noindent
(3)  The Neumann Green's function satisfies the differential equation
\begin{equation}
{\cal L}_{\br'}G_N(\br,\br')=\delta(\br-\br')-1/V,
\label{eq:g3ds}
\end{equation}
(4)  with the homogeneous boundary condition
\begin{equation}
{\bf\hat n'}\cdot\nabla G_N(\br,\br')=0
\label{eq:gbc3ds}
\end{equation}
on all bounding surfaces.\\
\\
\noindent 
(5) The solution to Eq.\ (\ref{eq:lp3ds}) is given by
\begin{equation}
\phi(\br)=<\phi>+\int G(\br,\br')\rho(\br')d\tau'-\int{\bf dS'}\cdot G(\br,\br')p(x')\nabla'\phi(\br'),
\label{eq:3dsltns}
\end{equation} 
(6)  where $<\phi>$ is an arbitrary constant that equals the average value of $\phi(\br)$ in the volume.\\
\\
\noindent
(7)  The Neumann Green's function is not necessarily symmetric, but can always be made symmetric by adding a function of {\br} to
$G_N(\br,\br')$.  Adding any function of $\br$ to a Neumann Green's function does not change its actions.

Of these seven differences, only number (1) is generally mentioned in Math Physics or EM textbooks.    Some books\cite{arf,but,jf} give the homogeneous boundary condition in Eq.\ (\ref{eq:gbc3ds}), but don't mention that it is inconsistent with omitting the $1/V$ term in  Eq.\ (\ref{eq:g3ds}).  Reference 4 recognizes this by making the Green's function boundary condition inhomogeneous, but this is not always a satisfactory remedy, as we have discussed above.  
%A Green's function with inhomogeneous boundary conditions can be just as hard to find as the original solution.  For instance, constructing a Green's function by an expansion %in eigenfunctions requires homogeneous boundary conditions.  


\begin{thebibliography}{9}
\bibitem{con}This constraint is noted on p.\ 66 of Ref.\ 3.
\end{thebibliography}
\end{document}